\documentclass[useAMS,usenatbib]{mn2e}
\usepackage[pdftex]{graphicx}     
\usepackage{epstopdf}
\usepackage{url}                  

\title[Brighter OAs at high $z$?]{Are GRB optical afterglows relatively brighter at high $z$?}

\author[A. Imerito, D. Coward, R. Burman and D. Blair]{A. Imerito$^{1}$\thanks{E-mail: alan@physics.uwa.edu.au}, D. M. Coward$^{1}$\thanks{E-mail: coward@physics.uwa.edu.au}, R. R. Burman$^{1}$ and D. G Blair$^{1}$\\
$^{1}$School of Physics, University of Western Australia, M013, Crawley WA 6009, Australia}

\begin{document}

\date{Accepted xxx. Received yyy; in original form zzz;    \textbf{~~~this draft: \today  ~~~Rev 38}}
\pagerange{\pageref{firstpage}--\pageref{lastpage}} \pubyear{2009}

\maketitle

\label{firstpage}

\begin{abstract}
The redshift distribution of gamma-ray bursts (GRBs) is strongly biased by selection effects. We investigate, via Monte Carlo simulations, one possible selection effect that may be modifying the {\it Swift} GRB redshift distribution. We show how telescope response times to acquire a GRB redshift may, via the Malmquist effect and GRB optical afterglow brightness distribution, introduce a bias into the average of the observed redshift distribution. It is difficult to reconcile a recently reported correlated trend between telescope response time and average redshifts unless we employ a redshift-dependent optical afterglow distribution. Simulations of this selection effect suggest that GRB optical afterglows may have been either intrinsically brighter early in the Universe or suffered less local host galaxy extinction.
\end{abstract}


\section{Introduction}

The NASA {\it Swift} satellite, launched in 2004 November, heralded a new era of rapid GRB localization. X-ray and UV telescopes on board {\it Swift} provided the means to localize GRBs with small error boxes, so that dedicated ground-based telescopes could image the fading optical afterglow (OA). Interestingly, only about 50\% of localized GRBs were identified with an optical afterglow prior to {\it Swift}. {\it Swift}'s sensitivity, combined with the growing number of rapid response ground-based telescopes capable of spectroscopy, promised to fill the gaps of GRB redshifts. Surprisingly, this did not happen: optical/NIR afterglows have been found for nearly $80\%$ of GRBs but only 40--50\% have measured redshifts \citep{Tanvir2007}. The synergy between Swift's sensitivity and localization capabilities and the growing number of rapid response ground-based telescopes has greatly improved the number of GRB redshifts that could be determined.

\section{GRB redshift selection effects}

The probability of GRB redshift measurement is proportional to the signal-to-noise ratio of the absorption or emission lines of the OA. Usually, multiple prominent lines are required, but this condition is hampered because GRB OA brightness decays roughly as $\sim 1/t$. In addition, many GRB host galaxies are too faint for redshifts to be obtained, so that the time taken to image and acquire spectra becomes critical. This was first pointed out by \citet{Fiore2007} for the observed discrepancy between the {\it HETE} and {\it BeppoSAX} redshift distributions compared to {\it Swift}. \citet{Coward2009} showed that the time taken to acquire a redshift measurement has led to a selection effect that is biasing the {\it Swift} redshift distribution. 


 \citet{Coward2007} and \citet{Coward2008} showed that for $z \sim 0-1$, the GRB redshift distribution should increase rapidly because of increasing differential volume sizes and strong SFR evolution. Until mid-2007, this characteristic in the {\it Swift} redshift distribution was not apparent. To account for this discrepancy, they argue that other biases, independent of the {\it Swift} sensitivity, are required. The lack of measured redshifts for $z \sim 1-2$ up to mid-2007, discussed by \citet{Coward2008}, is at least partially explained by selection effects from ground-based optical telescopes. 
 
\subsection{GRB redshifts and their statistics}

A shift of the mean of the GRB redshift distribution was observed in the early part of the {\it Swift} mission \citep{Berger05}. This was attributed to the improved sensitivity and more accurate localization by {\it Swift}, resulting in a bias for fainter and higher redshift bursts. \citet{Jakobsson2006} showed that within the first year of {\it Swift} the mean redshift for a subset of 28 bursts had increased to approximately 2.8, about double that of the pre-{\it Swift} average redshift. 

Assuming that satellite sensitivity is the dominant factor for determining redshift statistics, one could assume that the high mean redshift observed in the early part of the mission would remain fairly constant. If other factors impact on the statistics, such as the time taken to acquire high signal-to-noise spectra, the statistics may well reflect this. Assuming that {\it Swift}'s sensitivity and GRB localization ability has not degraded over time, one must consider the next link in the chain for measuring GRB redshifts: optical follow up by telescopes capable of spectroscopy. 



This basic idea can be related to GRB optical follow-up in terms of the efficiency of localizing, imaging and obtaining high quality spectra suitable for measuring redshift. In particular, a critical factor that determines the efficiency of these tasks is the time it takes to localize the rapidly fading OA and to obtain high signal-to-noise spectroscopy.

\begin{figure}
\includegraphics[scale=0.45]{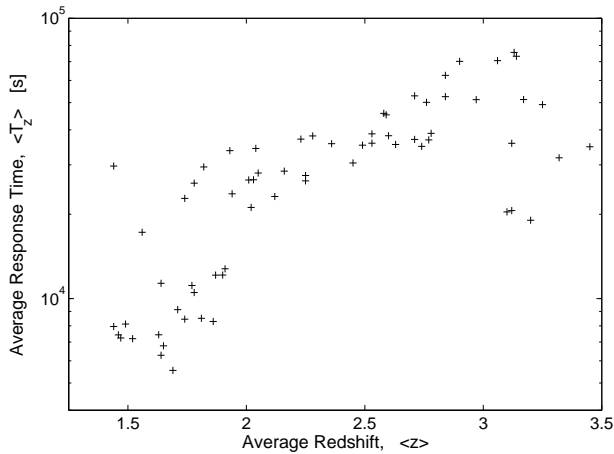}
\caption{The correlation of average response times, $\langle T_\mathrm{z}\rangle$, to acquire a spectroscopic redshift with the mean of the redshift distribution, $\langle z\rangle$ (adapted from fig. 3 of \citet{Coward2009}). The average redshifts were obtained from a moving average filter (sliding $z$-window) applied to the $z$ data. The average response times are calculated for each of the $z$-windows and plotted against the mid-point of their respective windows. The raw data response times are all measured from {\it Swift}'s BAT trigger to the time of obtaining a high-quality redshift spectrum.}
\label{fig:avTz_avz_Data}
\end{figure}

A downward drift in the {\it Swift} GRB mean redshift has been reported by \citet{Burrows2008}, and is reflected in the updated sample of \citet{Jakobsson2006} $\langle$\url{http://www.raunvis.hi.is/~pja/GRBsample.html}$\rangle$. The {\it Swift} team recently confirmed that this bias is not due to the satellite triggering algorithms \citep{Osborne2009}. Recently, \citet{Coward2009} has indicated that the source of this bias is most likely a selection effect from the improvement over time of the ground-based telescope efficiency to obtain spectroscopic redshifts. This results in decreasing telescope response times\footnote{Telescope `response time', $T_\mathrm{z}$, is defined here as the time from the start of the prompt $\gamma$-ray burst to the measurement of a redshift from the optical afterglow.}, $T_\mathrm{z}$. The intriguing result from this study is the decrease in average redshift, $\langle z\rangle$, of the bursts with decreasing average response time, $\langle T_\mathrm{z}\rangle$, illustrated in Fig. \ref{fig:avTz_avz_Data}. This is the opposite of what one would expect for a brightness-dependent bias. To understand this relationship one must consider how $\langle z\rangle$ is modified by the magnitude limit of a telescope and the brightness distribution of the OAs.

The observation by any magnitude-limited telescope of the OA events described aboved is subject to a familiar selection effect called the Malmquist bias \citep{Teerikorpi1997,  Butkevich2005}. This bias results from the preferential selection of intrinsically brighter members of a particular source luminosity distribution as their distance increases. This bias occurs whatever the luminosity distribution of the sources.

\section{Simulating a telescope-dependent redshift distribution}

The aim of this paper is restricted to identifying possible reasons for the positive correlated trend in the observed data shown in Fig. \ref{fig:avTz_avz_Data}. To do this we build a simple Monte Carlo model of OAs to study the effect of varying telescope response times. The model serves as an exploratory tool to enable a semi-quantitative study of the unexpected nature of the $\langle T_\mathrm{z}\rangle$--$\langle z\rangle$ relation reported above.


Two probability density functions (pdfs) are central to this model: the first is the OA luminosity distribution (i.e., normalized luminosity function, LF\footnote{LF refers to the luminosity function of the GRB optical afterglows.}), $\varphi(L)$,  and the second is the normalized (over $z$) volume distribution of GRB events, $\mathrm{d}R(z)/\mathrm{d}z$.

The spatial distribution is defined by the differential rate equation (see \citealt{Coward2007}; \citealt{Coward2001}):
\begin{equation}
\label{eqn:dR_dz}
\mathrm{d}R(z)/\mathrm{d}z = 4 \pi (c^3 r_0/H^3_0) e(z) F(z, \Omega_\mathrm{M} , \Omega_\mathrm{\Lambda})/(1+z)\;,
\end{equation}
where $\mathrm{d}R/\mathrm{d}z$ is the GRB differential event rate in units of $\mathrm s^{-1}$~per unit redshift, $e(z)$ is the dimensionless source rate density evolution function (scaled so that $e(0)=1$), $r_0$ is the GRB local rate density and $F(z, \Omega_\mathrm{M}, \Omega_\mathrm{\Lambda})$ is a dimensionless cosmology-dependent function (eqn. 13.61 of \citealt{Peebles1993}). A flat cosmology is assumed with $H_0$ = 71 km s$^{-1}$ Mpc$^{-1}$, $\Omega_M$ = 0.7 and $\Omega_\Lambda$ = 0.3. We assume that GRBs follow the star formation rate and take $e(z)$ to follow the SFR model of \citet{Hopkins2006}. The differential volume factor is the dominant factor in this expression.

For $\varphi(L)$, we used eqn. 1 of \citet{Johannesson2007},
\begin{equation}
\label{eqn:LF}
\varphi(L) = C\left(\frac{L}{L_0}\right)^{-\lambda}\exp\left(-\frac{\ln^2(L/L_0)}{2\sigma^2}\right)\exp\left(-\frac{L}{L_0}\right)\;,
\end{equation}
where $C$ is a normalization constant, $L_0$ a characteristic luminosity and $\sigma$ and $\lambda$ are parameters that control the shape of the function. Values for $\sigma$ and $\lambda$ were selected to match the R-band lightcurve from \citet{Johannesson2007} (see their fig. 1). This LF was converted into a pdf by normalizing it over the luminosity range for OAs (which, in absolute magnitude terms, we took as $[-25,-19]$ from fig. 3 of \citet{Kann2008} or fig. 5 of \citet{Kann2007}\footnote{These magnitudes are standardized to one day after the burst. For the simulation they were then scaled to the time of interest using the $t^{-\alpha}$ relationship (see Eqn. \ref{eqn:Pogson}).}). $L_0$ defines the upper exponential cutoff in the function and is set by the luminosity corresponding to an absolute magnitude of -25 (using, e.g., eqn. 2 of \citealt{Salpeter1986}\footnote{But note the typographical error in the equation of this reference: `$L$' is written in the denominator instead of `$L_{\sun}$'.}).

We also include in our simulations a redshift dependent LF\footnote{We use the term `evolving LF' to mean a luminosity function whose base luminosity increases `fast enough' with $z$ while preserving its functional shape. Note also that we are here using a brightening LF as a mathematical device. Physical interpretations are mentioned in Section 4.}. Two models are employed:
\begin{enumerate}
\item a continuously brightening luminosity distribution with $z$ (upper panel of Fig. \ref{fig:Mc_z}, where the brightening is from $z = z_\mathrm{evolv} = 1.5$ onwards); and
\item a discontinuous change (lower panel of Fig. \ref{fig:Mc_z}, where the brightening is from $z = z_\mathrm{evolv} = 4$ onwards).
\end{enumerate}
In both cases, modification of the LF merely involved shifting this distribution to higher brightness as $z$ increases without altering the shape of the LF itself.

To construct a synthetic population of afterglows an ensemble of GRB events is first generated in redshift space using Eqn. \ref{eqn:dR_dz}. Afterglow luminosity values ($M_\mathrm{c}$) are then generated according to the pdf of Eqn. \ref{eqn:LF} and randomly assigned to one of the previously created events in redshift space, thus producing a synthetic population of OAs. The OA luminosities are all assumed to decay as $(T_\mathrm{z} - t_0)^{-\alpha},$ with $\alpha = 1$ and $t_0$ the decay reference time (\citealt{Kobayashi2007}). With late-phase decay we can use the GRB response time, $T_\mathrm{z}$, to describe the decay by $T_\mathrm{z}^{-\alpha}$. 
(All times are measured in seconds in the observer's rest frame.) At specific times after the burst ($T_\mathrm{z,i},\;\mathrm{i} = 1,2,3,...$) we calculate the average redshift, $\langle z_\mathrm{i}\rangle$, for all the observable OAs. At these times, $t_\mathrm{i}\; (\equiv  T_\mathrm{z,i}$), after the start of the optical lightcurve decay, the apparent magnitudes $m(t_\mathrm{i})$ are calculated from the (temporally-modified) Pogson's formula:
\begin{equation}
\label{eqn:Pogson}
m(t_\mathrm{i}) = M_\mathrm{c} + 5\log_{10}(d_\mathrm{L}/10) + \frac{5\alpha}{2} \log_{10}(t_\mathrm{i}/t_\mathrm{c})\;.
\end{equation}
The calibration absolute magnitude for the OA is defined by $M_\mathrm{c}\equiv M(t_{\mathrm{c}})$, and $d_\mathrm{L}$ is the luminosity distance in parsecs, calculated via the analytical form given by \citet{Pen1999}. The calibration time for all afterglows, $t_\mathrm{c}$, is one day, as per \citet{Kann2008} or (2007).

A telescope limiting magnitude ($m_\mathrm{l}$) is used to define a brightness threshold for obtaining a redshift. Any simulated apparent magnitude dimmer than this threshold is not observable and, hence, no redshift is measured. So, for each $t_\mathrm{i}$ all events for which $m(t_\mathrm{i})\leq m_\mathrm{l}$ are taken to be observable. For this simulation we have taken $m_\mathrm{l} = 23$, so as to approximate a VLT-class instrument.

Fifty one Monte Carlo runs, each with a population of 5,000 OAs, were executed with each using different random number seeds to generate the OA populations. For each $t_\mathrm{i}$, the average redshift of all the observable OAs, $\langle z\rangle$, was calculated over all runs. The results were found to be insensitive to the seeds used.

The simulation makes several simplifying assumptions while still retaining the gross features of the OA distribution and decay:
\begin{enumerate}
\item We assume one limiting telescope magnitude, $m_\mathrm{l}$ --- i.e., a single telescope class (and identical seeing conditions) are assumed for OA spectroscopic observations.
\item The temporal decay is modelled by a single power law with the decay index, $\alpha$, taken to be the same for all OAs and all $z$.
\item We ignore any ($\gamma$-ray) triggering thresholds and field-of-view limitations of a satellite detector.
\end{enumerate}

\section{Results}

\begin{figure}
\includegraphics[scale=0.45]{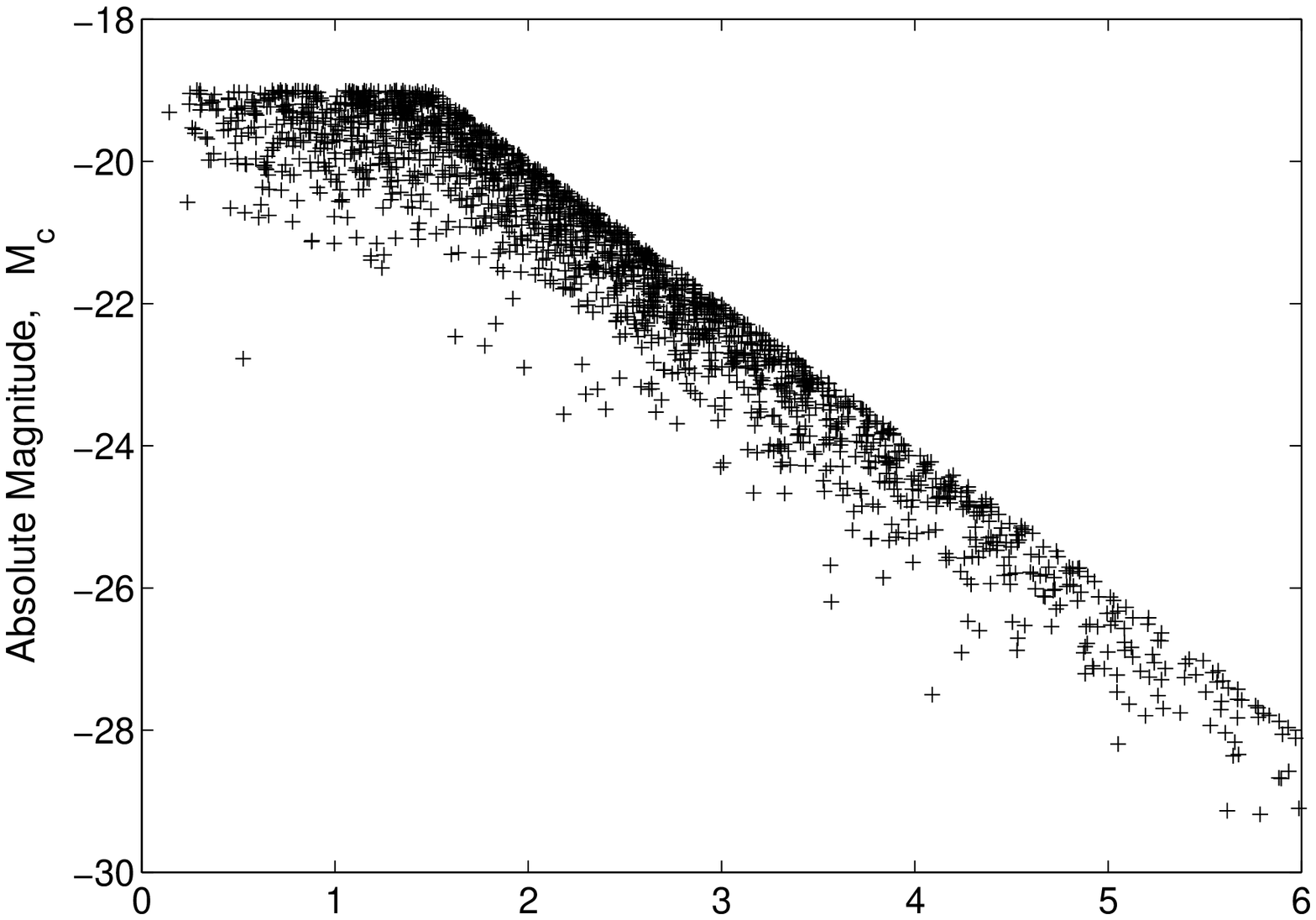}
\includegraphics[scale=0.45]{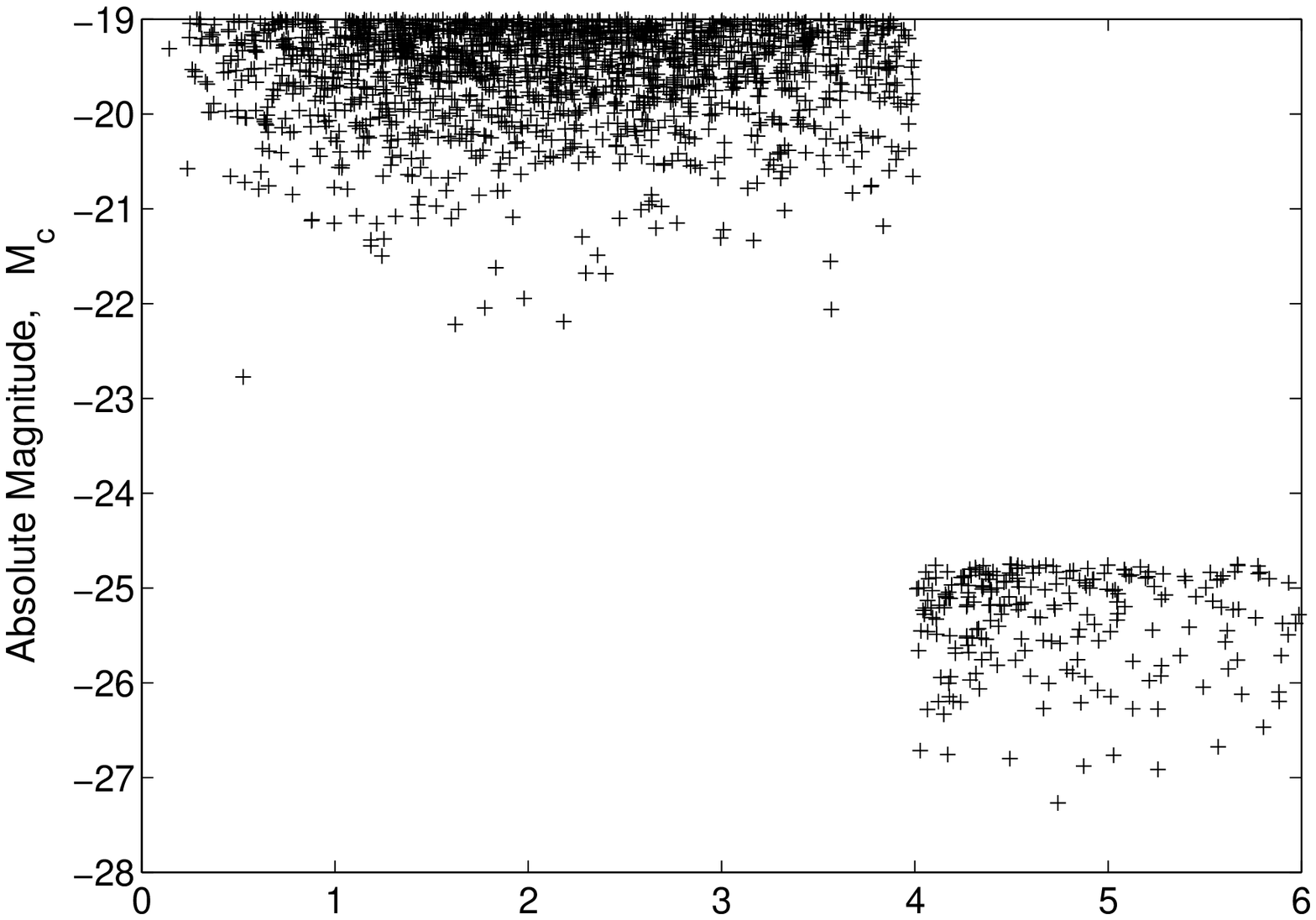}
\caption{Luminosity-redshift plots (\citet{Spaenhauer1978} diagrams) for two example evolving OA LFs at calibration time, $t_\mathrm{c}$. The upper panel trials a continuously brightening population starting from $z = 1.5$. The lower panel trials a discontinuous change, with brightening starting at $z = 4$. Each trial function above results in a growing $\langle z\rangle$ with $T_\mathrm{z}$. The purpose of using two different luminosity evolutions is merely to demonstrate qualitatively that enough brightness at high $z$ will result in $\langle z\rangle$ increasing with telescope response time, as exemplified by the simulation results in Fig. \ref{fig:Tz_Avz_Sim}.}
\label{fig:Mc_z}
\end{figure}

\begin{figure}
\includegraphics[scale=0.45]{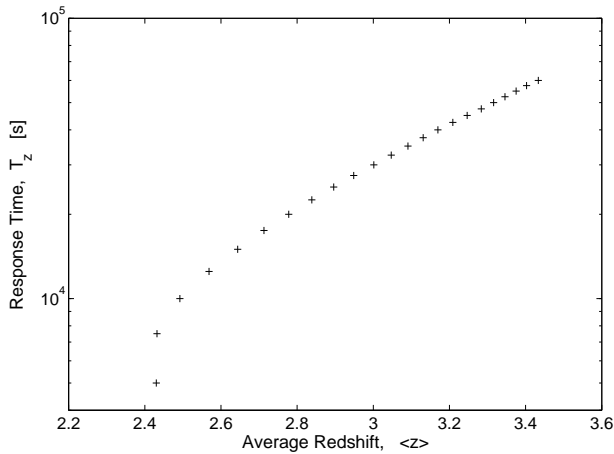}
\caption{The $z$-averaged results of all simulation runs showing the correlation of mean redshift with telescope response times to acquire a spectroscopic redshift. The LF evolution used in the simulation is that of the upper panel of Fig. \ref{fig:Mc_z}. The trend towards higher $\langle z\rangle$ with increasing $T_\mathrm{z}$ is the result of an OA population brightening with $z$, for $z > z_\mathrm{evolv}$ A similar result occurs for the two-component LF shown in the lower panel of Fig. \ref{fig:Mc_z}. Redshifts and response times obtained from the simulations are representative only, as they depend on various parameter values (e.g., telescope limiting magnitude $m_\mathrm{l}$, luminosity decay index $\alpha$) and the distribution of these parameters. Our model is an idealization which uses only a single value for each parameter. Additionally, one of the most influential factors affecting results is the function describing LF evolution --- something which is at present unknown.}
\label{fig:Tz_Avz_Sim}
\end{figure}

Using the same LF for all $z$ (i.e., a non-evolving LF), the simulated $T_\mathrm{z}$--$\langle z\rangle$ data consistently shows a negative correlation (i.e., the $T_\mathrm{z}$--$\langle z\rangle$ plot has a negative slope). This is expected from the Malmquist bias with a non-evolving LF, but it is opposite to the observational evidence reported by \citet{Coward2009}. By increasing the luminosity of the OA population with increasing redshift it is possible for our model to reproduce the trend of the observed data. 
We note that an evolving LF in this model may result from any of the following reasons, or a combination of them:
\begin{enumerate}
\item An intrinsic brightening of the OA population with increasing $z$ (e.g., fig. 2(c) of \citet{Badjin2009}).
\item Less local host extinction with increasing $z$.
\item Increasing energy shifted into the optical band from higher frequencies with increasing redshift (K-correction).
\end{enumerate}
The above will all manifest themselves as an increase in the overall brightness of the OAs, with or without the original source itself being intrinsically brighter. Hence, when modifying the OA LF in this simulation we understand that it may be due to a change in intrinsic source brightness, local extinction or frequency downshifting --- or a combination of these. So, with this understanding, whenever we speak of `brightening OAs' or `evolving LF' we mean an increase in the overall brightness of the OA at redshift $z$, whatever the cause.


Brightening the OAs with $z$ (via, e.g., either of the two trial luminosity distributions shown in Fig. \ref{fig:Mc_z}) yields a positive slope, similar to the observed data (compare Figs. \ref{fig:avTz_avz_Data} and \ref{fig:Tz_Avz_Sim}). With a monotonically brightening\footnote{The choice of a linearly brightening absolute magnitude with $z$ is taken purely on the grounds of simplicity as there is currently no evidence to favour any particular function.} LF with redshift (upper panel of Fig. \ref{fig:Mc_z}), the simulated $T_\mathrm{z}$--$\langle z\rangle$ data yields a similar positive trend to the observations (compare Figs. \ref{fig:avTz_avz_Data} and \ref{fig:Tz_Avz_Sim}). Similar results hold if we use a two-component model, such as that displayed in the lower panel of Fig. \ref{fig:Mc_z}.

The primary factor in achieving the result we seek is an increase in the relative luminosity of OAs with redshift.

If we are given the form of an evolving LF then the redshift where the LF starts evolving, $z_\mathrm{evolv}$, can be approximated. For a linearly brightening OA (upper panel of Fig. \ref{fig:Mc_z}), setting $z_\mathrm{evolv}$ too high (from Eqn. \ref{eqn:dR_dz}, the GRB event rate peaks at $z\approx 2$) will result in too few OAs in the more distant Universe ($z > z_\mathrm{evolv}$) with the consequence that the $T_\mathrm{z}$--$\langle z\rangle$ relation weakens and tends to a negative correlation. There is also a trade-off between the rate of brightening and $z_\mathrm{evolv}$: the faster the rate of brightening with $z$, the smaller $z_\mathrm{evolv}$ can be while still producing a positive correlation for $T_\mathrm{z}$--$\langle z\rangle$.

It is instructive to note that the direction of variation of $\langle z\rangle$ differs depending on whether or not we are dealing with an evolving or non-evolving LF. The effect on $\langle z\rangle$ of individually varying $\alpha$, $m_\mathrm{l}$ and $T_\mathrm{z}$ is summarized in Table \ref{tbl:av_z}.

In summary, we find that with a fast enough brightening of OA luminosity with $z$, the average GRB redshift distribution of the observed afterglows increases with telescope response time, in accord with observations. If a non-evolving LF is employed the $T_\mathrm{z}$--$\langle z\rangle$ relation obtained from our simulation is in contradiction to the data. This result is insensitive to the shape of the LF but is dependent upon the function defining the evolution of the LF with $z$.

\begin{table} 
\caption{The direction of change of $\langle z\rangle$ for a `fast enough' evolving LF. A hyphen in the table indicates a constant value for the parameter, $\uparrow$ indicates an increase and $\downarrow$ a decrease in value. The results are insensitive to the LF itself but depend on the rate of LF brightening with $z$.}
\label{tbl:av_z}
\begin{tabular}{ccccc}
\hline
   &   &    & {\bf Non-Evolving LF} &   {\bf Evolving LF}                             \\
 {\bf $\alpha$}  &  {\bf $m_\mathrm{l}$} &   {\bf $T_\mathrm{z}$}  &  {\bf $\langle z\rangle$}  &  {\bf $\langle z\rangle$}   \\\hline
$\uparrow$ & - & -               &  $\downarrow$       &  $\uparrow$                  \\
- & $\uparrow$ & -               &  $\uparrow$         &  $\downarrow$                \\
- & - & $\uparrow$               &  $\downarrow$       &  $\uparrow$                  \\
\hline
\end{tabular}
\end{table}


\section{Discussion}

As a result of brightening the GRB OA LF with $z$, the proportion of bright OAs will increase with $z$. Thus, if the LF brightens fast enough with $z$ it can more than compensate for the the deceasing number of OAs after $z\approx 2$ and the dimming effect of distance. In particular, for a long response time the very bright and distant OAs increase the $\langle z\rangle$ as they start their decay from a much higher luminosity than closer and relatively dimmer OAs. Hence, because dimmer OAs fade out quicker as $T_\mathrm{z}$ increases, the high-$z$ OAs are observable for longer despite being more distant. This results in an increasing $\langle z\rangle$ as the response time increases.

Increasing the number of OAs at high $z$ \citep{Daigne2006} with an unevolved LF did not produce a positive correlation in $T_\mathrm{z}$--$\langle z\rangle$ in our simulation and neither did modifying the shape of the LF. In fact, we found that the $T_\mathrm{z}$--$\langle z\rangle$ result was insensitive to the shape of the OA LF but determined by the rate of the LF evolution.

Other parameters affect the results in a mostly quantitative manner: the more sensitive the telescope the stronger the $T_\mathrm{z}$--$\langle z\rangle$ correlation becomes. Conversely, a larger $\alpha$ results in a weaker correlation.

Within the limitations of our model, we find evidence that the observed time-dependent selection effect may be the result of an  evolving LF (smooth or not) for optical afterglows, with the relatively brighter afterglows occuring in the early Universe. The weak positive correlation between OA luminosity and $E_{\mathrm{\gamma,iso}}$ reported by \citet{Kann2007} and suggested by fig. 2(a) of \citet{Liang2006} raises the possibility that GRBs may have been evolving through time, with $\gamma$-ray bursts being more energetic in the early Universe.

Our use of an evolving LF in simulations is consistent with work that finds GRB bursts to be distributed between near low-luminosity and more distant and luminous GRBs (e.g., \citealt{Liang2007, Cobb2006}, \citealt{Chapman2007}). \citet{Nardini2008}, \citet{Kann2007}, \citet{Liang2006} and the left-hand panel of fig. 7 of \citet{Melandri2008}\footnote{Although these authors argue that their OA data does not support bimodality, the lack of nearby luminous bursts in their fig. 7 tends to support it.} suggest that OA luminosities themselves may be partitioned into two or three populations.

In summary, we find that the trend of $T_\mathrm{z}$--$\langle z\rangle$ is independent of the LF but governed by the strength of the LF evolution and the redshift regime where this evolution takes place.\\

On a more abstract level, we note that attempts are often made to eliminate or work around selection effects. However, in this study we have found that a time-dependent selection effect has been helpful in extracting new insight into the GRB OA distribution. Perhaps in other studies, selection effects may also provide more illumination than obscuration.

\section {Acknowledgments}

We thank the referee for his/her diligent work and for providing several suggestions that have helped to clarify the results. AI is supported by the Australian Research Council (ARC) grant LP0667494. DMC is supported by ARC grants DP0877550, LP0667494 and the University of Western Australia.

\label{lastpage}

\bsp

\end{document}